\documentclass[11pt,final,journal,letter,oneside,twocolumn]{IEEEtran}

\usepackage{multirow}
\usepackage{tabulary}
\usepackage{srcltx}
\usepackage{psfrag}
\usepackage{epsfig}
\usepackage{color}
\usepackage[tbtags,sumlimits,nointlimits,reqno]{amsmath}
\usepackage{amssymb}
\usepackage{showkeys}   
\usepackage{subfig}

\newcommand{\be}{\begin{eqnarray}}
\newcommand{\ee}{\end{eqnarray}}

\graphicspath{{./fig/}}

\newcommand{\figref}[1]{\text{Fig.~\ref{#1}}}

\begin{document}
%
\title{%
Effect of Spatial Dispersion on Surface Waves Propagating Along Graphene Sheets}
%
%
%
%
%
\author{%
       J.~S.~Gomez-Diaz,~J.~R.~Mosig, and J.~Perruisseau-Carrier
\thanks{%
        J. S. Gomez-Diaz and J. Perruisseau-Carrier are with the Adaptive MicroNanoWave Systems, LEMA/Nanolab,
        \'Ecole Polytechnique F\'ed\'erale de Lausanne.
        1015 Lausanne, Switzerland.
        E-mail: juan-sebastian.gomez@epfl.ch, julien.perruisseau-carrier@epfl.ch}
\thanks{%
        J. Mosig is with the Electromagnetics and Acoustics Laboratory (LEMA),
        \'Ecole Polytechnique F\'ed\'erale de Lausanne.
        1015 Lausanne, Switzerland.
        E-mail: juan.mosig@epfl.ch}}
%
\maketitle
\begin{abstract}
We investigate the propagation of surface waves along a spatially dispersive graphene sheet, including substrate effects. The proposed analysis derives the admittances of an equivalent circuit of graphene able to handle spatial dispersion, using a non-local model of graphene conductivity. Similar to frequency selective surfaces, the analytical admittances depend on the propagation constant of the waves traveling along the sheet. Dispersion relations for the supported TE and TM modes are then obtained by applying a transverse resonance equation. Application of the method demonstrates that spatial dispersion can dramatically affect the propagation of surface plasmons, notably modifying their mode confinement and increasing losses, even at frequencies where intraband transitions are the dominant contribution to graphene conductivity. These results show the need for correctly assessing spatial dispersion effects in the development of plasmonic devices at the low THz band.
\end{abstract}

%
\IEEEpeerreviewmaketitle


\section{Introduction}
\label{Introduction}
The propagation of electromagnetic waves along graphene sheets have recently attracted significant attention \cite{Hanson08}, \cite{Hanson09}, \cite{Jablan09}. Graphene, thanks to its interesting electrical and optical properties \cite{Novoselov04}, provides new possibilities for surface wave propagation at optical, infrared, and low THz frequencies \cite{Jablan09}, \cite{Engheta11}. Specifically, graphene enables the development of novel plasmonic devices at relatively low frequencies \cite{Jablan09}, \cite{Barnes03}, \cite{Bludov10}, \cite{Tamagnone12_apl} in contrast with noble metals which only allows plasmonic propagation in the visible range \cite{Pitarke07, Wang11}. In addition, graphene is inherently tunable, via the application of a electrostatic or magnetostatic bias field, leading to novel reconfiguration possibilities \cite{Engheta11}, \cite{Gusynin2007}. Many research groups have theoretically studied the propagation of plasmons (i.e., electromagnetic waves propagating at the interface between a conductor and a dielectric) along graphene \cite{Hanson08,Hanson09,Jablan09,Hwang07,sebas12_jap2}. Also, different configurations have recently been proposed to improve the characteristics of this propagation, including parallel plate pairs \cite{Hwang07, Hanson08_PPW} or waveguides \cite{Christensen11}. Furthermore, recent studies have demonstrated enhanced transmission through a stack of monolayer graphene sheets \cite{Kaipa12}. However, though graphene is known to be spatially dispersive \cite{Falkovsky07b}, spatial dispersion has usually been neglected in works related to wave propagation at low THz frequencies. In \cite{Hanson09}, a non-local model of the graphene tensor conductivity, valid in the frequency range where intraband contributions of graphene dominate, was proposed. Based on this model, \cite{Hanson09} also predicted that spatial dispersion would become significant for extremely slow surface waves. In case of graphene in free space, this occurs when interband contributions of graphene dominate (moderate to high THz frequencies). This model was also recently employed in \cite{Lovat13_Spatial_Dispersion} to study spatial dispersion effects in graphene-based parallel-plate waveguides, focusing on the particular case of graphene with uniform zero chemical potential.

In this work, we investigate the propagation of surface waves along a spatially dispersive graphene sheet, including substrate effects, and we show that spatial dispersion affects the behavior of these waves even in the low THz range. The sheet is characterized using the graphene tensorial conductivity obtained by applying the non-local spatially dispersive model of graphene presented in \cite{Hanson09}, \cite{Lovat13_Eucap}, \cite{Lovat_Hanson_13}, valid in the absence of external magnetostatic bias field, at frequencies where intraband contributions of graphene dominate, and for any value of graphene chemical potential. This model has been derived from the semi-classical Boltzmann transport equation assuming a linear electron dispersion, and uses the relaxation-time (RTA, see \cite{Dressel02}) and the so-called $k_\rho$-low approximations. Therefore, it is accurate in the case of relatively low values of the propagation constant $k_\rho$, while provides approximate results for moderate values of $k_\rho$. Here we employ this model instead of the general spatially-dispersive graphene conductivity model described in \cite{Falkovsky07b} because it allows to obtain closed-form expressions for the propagation constant of spatially-dispersive surface waves propagating on graphene sheets, thereby providing physical insight into their properties. The operator components of the conductivity tensor are then mapped onto the admittances of a rigorous Green's function-based equivalent circuit of graphene \cite{Lovat12}. These equivalent admittances, which are similar to those found in the analysis of frequency selective surfaces \cite{Munk00}, \cite{Maci05}, analytically show that the influence of spatial dispersion directly depends on the square of the wave propagation constant ($k_\rho$). Then, a transverse resonance equation (TRE) \cite{Collin69} is imposed to compute the dispersion relation of surface waves along spatially dispersive graphene. Similar to the case of non spatially dispersive graphene sheets, transverse electric (TE) and transverse magnetic (TM) modes are supported. Analytical dispersion relations are provided for TM and TE surface waves propagating along a graphene sheet embedded into an homogeneous medium, while approximate expressions are given for the case of graphene surrounded by two different dielectrics. The derived dispersion relations analytically show that the permittivity of the surrounding media, operation frequency and spatial dispersion similarly contribute to determine the characteristics of $k_\rho$, suggesting that the influence of graphene spatial dispersion in the propagating waves may be strongly affected by the environment of the sheet. Numerical results confirm that spatial dispersion is an important mechanism for wave propagation along graphene sheets, leading to surface modes that can significantly differ from those found neglecting spatial dispersion. These features, which include variations in the mode confinement and higher losses, should be rigorously taken into account in the development of novel plasmonic devices at the low THz band.

The paper is organized as follows. Section \ref{sec:equivalent_circuit_section} derives the analytical relations between the admittances of a Green's function-based equivalent circuit of graphene and the components of the spatially dispersive conductivity tensor. Then, Section \ref{sec:Dispersion_relationship} computes the dispersion relation of surface waves propagating along graphene, providing analytical expressions for the case of a graphene sheet embedded into an homogeneous media. Section \ref{sec:Numerical_results} discusses the characteristics of surface waves along spatially dispersive graphene, taking into account the surrounding media. Finally, conclusions and remarks are provided in Section \ref{sec:Conclusions}.
\section{Equivalent Circuit of a \\Spatially Dispersive Graphene Sheet}
\label{sec:equivalent_circuit_section}
\begin{figure} \centering
\subfloat[]{\label{fig:_Draw}
\includegraphics[width=0.7\columnwidth]{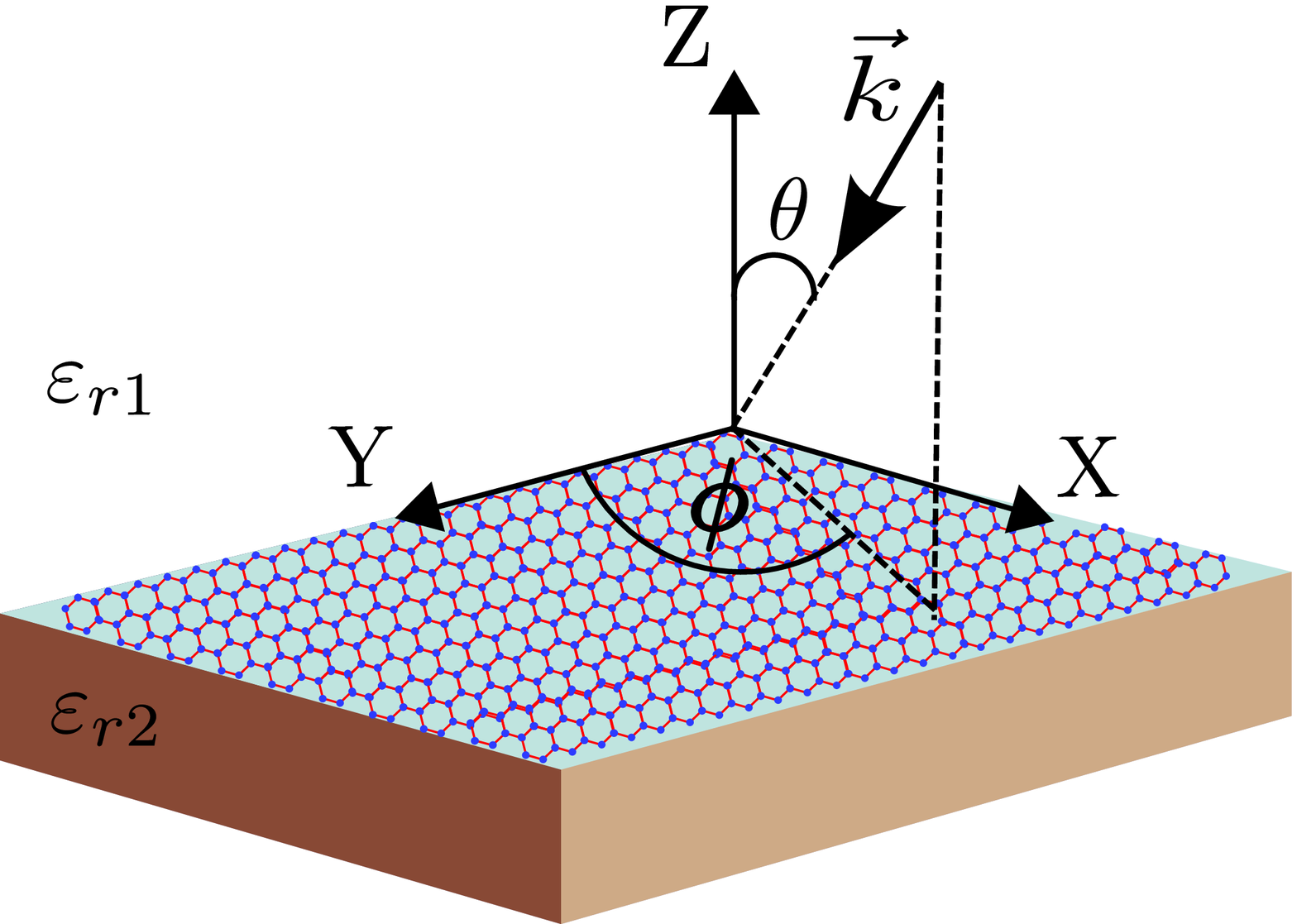}} \\
\subfloat[]{\label{fig:_Circuit_model_incidence}
\includegraphics[width=0.8\columnwidth]{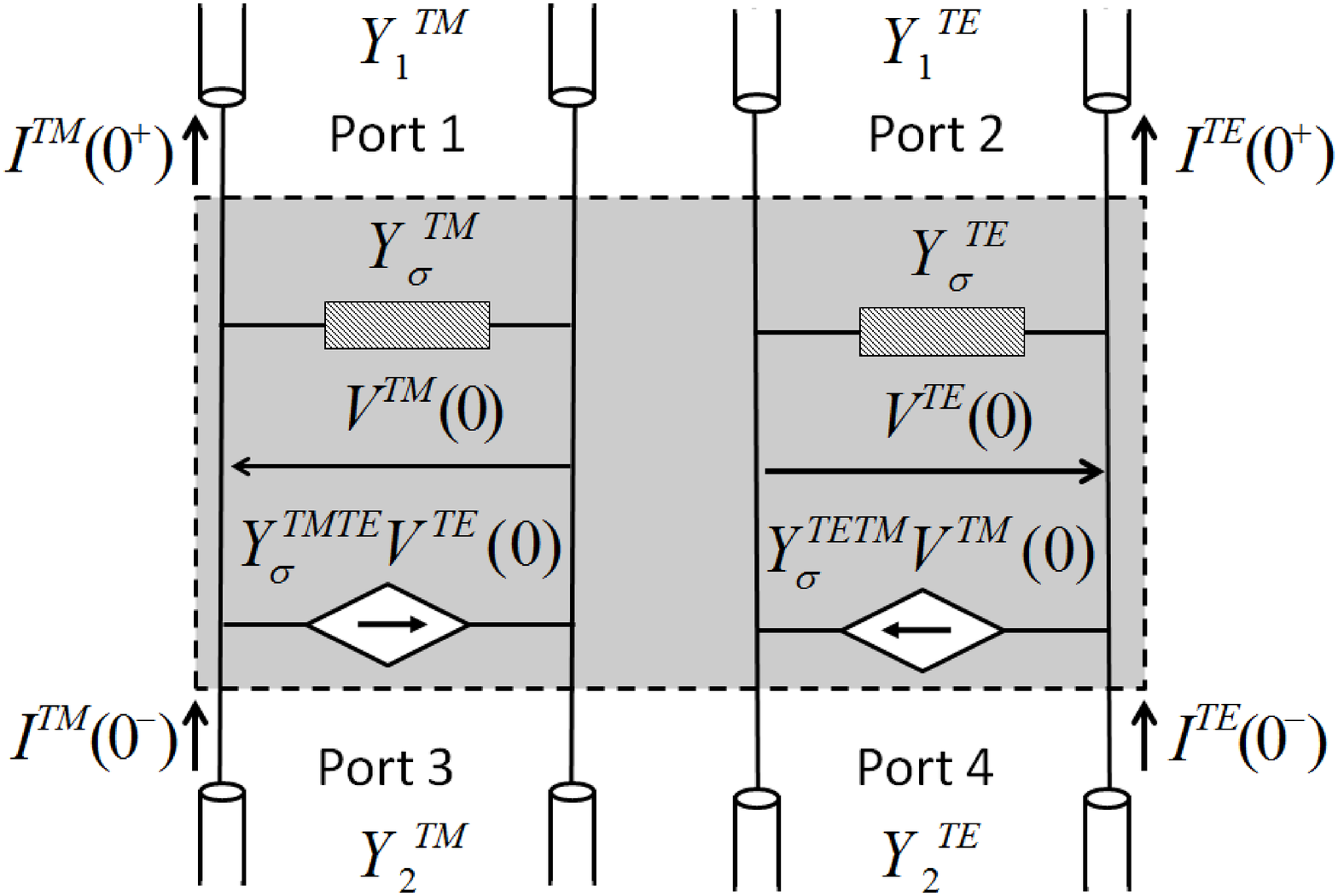}} \\
\caption{Spatially dispersive graphene sheet on a dielectric substrate under arbitrary plane-wave incidence (a) and its equivalent transmission line model \cite{Lovat12} (b).} \label{fig:_Main_scheme}
\end{figure}
Let us consider an infinitesimally thin graphene sheet in the plane $z=0$ and separating two media, as illustrated in Fig. \ref{fig:_Draw}. The sheet is characterized by the conductivity tensor $\overline{\overline{\sigma}}$, obtained by applying a spatially dispersive (non-local) model of graphene \cite{Hanson09} in the absence of external magnetostatic biasing fields ($\vec{B_0}=0$). This anisotropic conductivity reads
\begin{equation}
\overline{\overline{\sigma}}(\omega,\mu_c,\tau,T)=\left(\begin{array}{cc} \sigma_{x'x'} & \sigma_{x'y'} \\
                                                                    \sigma_{y'x'} & \sigma_{y'y'} \end{array} \right),
\label{eq:Conductivity}
\end{equation}
where $T$ is temperature, $\tau$ is the electron relaxation time, $\mu_c$ is the chemical potential and $\omega$ is the angular frequency. Due to the spatial dispersion of graphene, the conductivity components become operators \cite{Hanson09}, \cite{Hanson12_private}
\begin{equation}
\sigma_{x'x'}=\sigma_{lo}+\alpha_{sd}\frac{\partial^2}{\partial x'^2}+\beta_{sd}\frac{\partial^2}{\partial y'^2}, \label{eq:Conductivity_xx}
\end{equation}
\begin{equation}
\sigma_{y'y'}=\sigma_{lo}+\beta_{sd}\frac{\partial^2}{\partial x'^2}+\alpha_{sd}\frac{\partial^2}{\partial y'^2}, \label{eq:Conductivity_yy}
\end{equation}
\begin{equation}
\sigma_{x'y'}=\sigma_{y'x'}=2\beta_{sd}\frac{\partial^2}{\partial x'\partial y'}, \label{eq:Conductivity_cross}
\end{equation}
where
\begin{align}
\sigma_{lo}&=\frac{-jq_e^2k_BT}{\pi\hbar^2(\omega-j\tau^{-1})}\ln\left[2\left(1+\cosh\left(\frac{\mu_c}{k_BT}\right)\right)\right], \\
\alpha_{sd}&=\frac{-3v_F^2\sigma_{lo}}{4(\omega-j\tau^{-1})^2}, \quad
\beta_{sd}=\frac{\alpha_{sd}}{3},
\label{eq: Conductivity_support}
\end{align}
$k_B$ is the Boltzmann constant, $q_e$ is the electron charge, $v_F$ is the Fermi velocity ($\approx10^6$~m/s in graphene), and the subscripts ``lo" and ``sd" have been included to denote local and spatially-dispersive (non-local) terms. Note the presence of a minus sign in $\alpha_{sd}$, which was not included in \cite{Hanson09} due to a typo \cite{Hanson12_private}. It is worth mentioning that this model is insensitive to the orientation of the graphene lattice, and that anisotropy arises as the response of the material to the excitation, as usually occurs with spatial dispersion \cite{Hanson09}, \cite{Hanson12_private}. Consequently, the coordinate system employed in the anisotropic conductivity [$x'-y'$, see Eqs.~(\ref{eq:Conductivity_xx})-(\ref{eq:Conductivity_cross})] is related to the excitation, i.e. $\sigma_{x'x'}$ and $\sigma_{y'y'}$ provides the response parallel to the $E_{x'}$ and $E_{y'}$ fields, being $x'$ and $y'$ arbitrary perpendicular directions in the infinite sheet. Importantly, Eqs.~(\ref{eq:Conductivity_xx})-(\ref{eq: Conductivity_support}) are only strictly valid at frequencies where the intraband contributions of graphene conductivity dominates, usually the low THz regime. Intraband contributions correspond to electron transitions between different energy levels in the same band (valence or conduction), while interband contributions are related to electron transitions between different bands. The latter phenomenon becomes significant as frequency increases due to the higher photon energy \cite{Dressel02}.

A rigorous equivalent circuit of an anisotropic graphene sheet sandwiched between two media is shown in Fig. \ref{fig:_Circuit_model_incidence} \cite{Lovat12}. This four-port circuit relates input and output TE and TM waves through equivalent shunt admittances ($Y_\sigma^{TE}$ and $Y_\sigma^{TM}$, respectively), and the cross-coupling between the two polarizations through voltage-controlled current generators,  with coefficients $Y_\sigma^{TE/TM}$ and $Y_\sigma^{TM/TE}$ for TE-TM and TM-TE coupling, respectively. The relation between the anisotropic conductivity of graphene and these admittances was provided in \cite{Lovat12} for the case of a local model of graphene, i.e. when anisotropy is due to an external magnetostatic bias field. However, spatial dispersion effects have not been considered so far.

Here, we analytically obtain the admittances of the equivalent circuit shown in Fig. \ref{fig:_Circuit_model_incidence} using a non-local (spatially dispersive) model of graphene. For this purpose, we include the operator components of the tensor conductivity, Eqs. (\ref{eq:Conductivity_xx})-(\ref{eq:Conductivity_cross}), into the transmission-line network formalism for computing dyadic Green's functions in stratified media \cite{Lovat12, michalski97}. For the admittances derivation, we consider a uniform plane wave impinging on the graphene sheet from an arbitrary direction ($\theta$, $\phi$) of medium $1$ (see \figref{fig:_Draw}), and then we use the corresponding auxiliary coordinate system for the conductivity tensor. The incoming wave has a wavenumber $\vec{k}=k_x\hat{e}_x+k_y\hat{e}_y+k_z\hat{e}_z$, where $k_x=k_1\cos(\phi)\sin(\theta)$, $k_y=k_1\sin(\phi)\sin(\theta)$, $k_z=k_1\cos(\theta)$, $\hat{e}_x,\hat{e}_y,\hat{e}_z$ are unit vectors, and $k_0=2\pi/\lambda_0$ and $k_1=\sqrt{\varepsilon_{r1}}k_0$ are the wavenumbers of free space and medium $1$, respectively. Imposing boundary conditions in the graphene plane and following the approach described in \cite{Lovat12}, \cite{michalski97} these admittances can be written as
\begin{align}
Y_\sigma^{TE}(k_\rho)&=\sigma_{lo}+k_\rho^2[\alpha_{sd}+\beta_{sd}], \label{eq:Y_TE} \\
Y_\sigma^{TM}(k_\rho)&=\sigma_{lo}+k_\rho^2[\alpha_{sd}+\beta_{sd}], \label{eq:Y_TM} \\
Y_\sigma^{TE/TM}(k_\rho)&=Y_\sigma^{TM/TE}(k_\rho)=0, \label{eq:Y_cross}
\end{align}
where $k_\rho^2=k_x^2+k_y^2$ is the propagation constant of the wave traveling along the graphene sheet.

The importance of Eqs.~(\ref{eq:Y_TE})-(\ref{eq:Y_cross}) is two fold. First, they analytically indicate that the influence of spatial dispersion is directly proportional to $k_\rho^2$. Consequently, this phenomenon will be more important for very slow waves ($k_\rho\gg k_0$), which usually appear at moderately high THz frequencies (where interband contributions of graphene dominate and Eqs. (\ref{eq:Conductivity_xx})-(\ref{eq:Conductivity_cross}) are not strictly valid \cite{Hanson09}). However, note that slow waves can also be obtained in graphene sheets at low THz frequencies by using substrates with high permittivity constant \cite{Jablan09}. Second, the equivalent admittances do not depend on the direction of propagation of the wave along the sheet ($\phi$) and there is no coupling between the TM and TE modes. This is in agreement with the non-local model of graphene employed in this development \cite{Hanson09}, \cite{Hanson12_private} which assumes an isotropic infinite surface where spatial dispersion arises as a response to a given excitation.
\section{Dispersion Relation for a \\ Spatially Dispersive Graphene Sheet}
\label{sec:Dispersion_relationship}
The dispersion relation of surface waves on a spatially dispersive graphene sheet sandwiched between two different media can be obtained by imposing a transverse resonance equation \cite{Collin69} to the equivalent circuit shown in \figref{fig:_Circuit_model_incidence}. The solution of the TRE provides the propagation constant $k_\rho$ of the surface wave propagating along the sheet. Following the approach described in \cite{sebas12_jap2}, the desired dispersion relation can be obtained as
\begin{align}
\left(Y_1^{TE}+Y_2^{TE}+Y_\sigma^{TE}\right)&\left(Y_1^{TM}+Y_2^{TM}+Y_\sigma^{TM}\right)=0,
\label{eq:Dispersion_relation_general}
\end{align}
where
\begin{align}
Y_1^{TE}&=\frac{k_{z_1}}{\omega\mu_0}, \quad
Y_2^{TE}=\frac{k_{z_2}}{\omega\mu_0}, \quad
Y_1^{TM}=\frac{\omega\varepsilon_{r_1}\varepsilon_0}{k_{z_1}}, \nonumber \\
Y_2^{TM}&=\frac{\omega\varepsilon_{r_2}\varepsilon_0}{k_{z_2}},
k_{z_1}=\pm\sqrt{k_1^2-k_\rho^2},
k_{z_2}=\pm\sqrt{k_2^2-k_\rho^2}, \nonumber \\
k_1&=\sqrt{\varepsilon_{r_1}}k_0, \quad \text{and} \quad
k_2=\sqrt{\varepsilon_{r_2}}k_0
\label{eq:Media_definition}
\end{align}
are the TE and TM admittances, transverse propagation constant and wavenumber of media $1$ and $2$, respectively.

Eq. (\ref{eq:Dispersion_relation_general}) does not generally admit any analytical solution and must be solved using numerical methods \cite{Press96b}. Importantly, the mathematical solutions of this equation must be carefully checked to verify that they correspond to physical modes. Specifically, physical modes must fulfill the law of energy conservation, i.e. surface waves cannot be amplified while propagating along the structure, and the Sommerfeld boundary radiation condition \cite{Collin69}. Also, note that the solution of Eq.~(\ref{eq:Dispersion_relation_general}) leads to the propagation constant of a transverse electric (TE) or a transverse magnetic (TM) mode, as in the case of isotropic graphene \cite{Hanson08}. These cases are examined below.
\subsection{TM modes}
\label{sec:TM_modes}
The dispersion relation for TM modes propagating along a spatially dispersive graphene sheet can be obtained by solving $Y_1^{TM}+Y_2^{TM}+Y_\sigma^{TM}=0$ [see Eq. (\ref{eq:Dispersion_relation_general})]. This equation may be expressed as
\begin{align}
\frac{\omega\varepsilon_{r1}\varepsilon_0}{\sqrt{\varepsilon_{r1}k_0^2-k_\rho^2}}+\frac{\omega\varepsilon_{r2}\varepsilon_0}{\sqrt{\varepsilon_{r2}k_0^2-k_\rho^2}}=-\left[\sigma_{lo}+k_\rho^2(\alpha_{sd}+\beta_{sd})\right].
\label{eq:Dispersion_relation_TM}
\end{align}

If the graphene sheet is embedded into an homogeneous host medium, i.e. $\varepsilon_{r1}=\varepsilon_{r2}=\varepsilon_r$, the dispersion relation can be simplified to
\begin{align}
&k_\rho^6+k_\rho^4\left[\frac{2\sigma_{lo}}{\alpha_{sd}+\beta_{sd}}-\varepsilon_rk_0^2\right]-\nonumber \\ &k_\rho^2\left[\frac{2\varepsilon_r\sigma_{lo}k_0^2(\alpha_{sd}+\beta_{sd})-\sigma_{lo}^2}{(\alpha_{sd}+\beta_{sd})^2}\right]+\frac{4\omega^ 2\varepsilon_r^2\varepsilon_0^2-\varepsilon_rk_0^2\sigma_{lo}^2}{(\alpha_{sd}+\beta_{sd})^2}=0,
\label{Dispersion_relation_TM_homogeneous_medium}
\end{align}
which has $6$ complex roots but can be solved analytically using standard techniques \cite{Pipes71}. Also, considering two different media and assuming the usual non-retarded regime ($k_\rho\gg k_0$), Eq.~(\ref{eq:Dispersion_relation_TM}) reduces to
\begin{align}
k_\rho^3+k_\rho\frac{\sigma_{lo}}{\alpha_{sd}+\beta_{sd}}-j\omega\frac{\varepsilon_0(\varepsilon_{r1}+\varepsilon_{r2})}{\alpha_{sd}+\beta_{sd}}=0,
\label{eq:Dispersion_relation_explicit}
\end{align}
which is a cubic equation with three complex roots. Note that the operation frequency ($\omega$) and the permittivity of the surrounding media ($\varepsilon_{r1}$ and $\varepsilon_{r2}$) only appear in the free term of Eq.~(\ref{eq:Dispersion_relation_explicit}), where they multiply each other. Therefore, the behavior of surface waves propagating along a graphene sheet is mainly determined by this product, suggesting that similar responses will be obtained for larger frequencies if the permittivity is simultaneously reduced, or vice-versa [Similar conclusions are reached by closely examining Eq.~(\ref{Dispersion_relation_TM_homogeneous_medium})]. However, this does not mean that frequency and permittivity are fully interchangeable  since graphene conductivity itself is frequency-dependent. In the asymptotic limit, $\omega\varepsilon_0(\varepsilon_{r1}+\varepsilon_{r2})>>k_\rho\sigma_{lo}$, Eq. (\ref{eq:Dispersion_relation_explicit}) can be solved analytically and yields
\begin{align}
k_\rho\approx \sqrt[3]{j\omega\varepsilon_0 \frac{(\varepsilon_{r1}+\varepsilon_{r2})}{\alpha_{sd}+\beta_{sd}}}.
\label{eq:Dispersion_relation_explicit_asymptotic}
\end{align}
This equation shows that in the limit of very high frequencies, or a large permittivity of the surrounding media,
spatial dispersion will be the main phenomenon governing wave propagation. However, note that interband contributions of graphene conductivity, which are dominant at such high frequencies, have not been considered in this analysis.

In the absence of spatial dispersion ($\alpha_{sd}=\beta_{sd}=0$), Eq. (\ref{eq:Dispersion_relation_explicit}) simplifies to the usual expression \cite{Jablan09}
\begin{align}
k_\rho\approx j\omega\varepsilon_0 \frac{(\varepsilon_{r1}+\varepsilon_{r2})}{\sigma_{lo}}.
\label{eq:Dispersion_relation_explicit_NO_Spatial_Dispersion}
\end{align}

The cut-off frequency of the propagating TM modes can be obtained by numerically finding the lowest frequencies which fulfill Eq.~(\ref{eq:Dispersion_relation_TM}). A very good approximation is obtained by identifying the frequency range of $Im(\sigma_{lo})<0$, which is the condition for TM mode propagation along a non-spatially dispersive graphene sheet \cite{Hanson08}. Note that the cut-off frequency of the supported TM modes mainly depends on the characteristics of the local conductivity of graphene, which can be externally controlled by tuning the chemical potential of graphene via the field effect.
\begin{figure*}[!bth]
\begin{align}
\label{Dispersion_relation_TE_mode}
k_\rho=\pm\sqrt{\frac{-\sigma_{lo}\omega^2\mu_0^2(\alpha_{sd}+\beta_{sd})+2(1\pm\sqrt{1+\omega^2\mu_0^2(\alpha_{sd}+\beta_{sd})[\sigma_d+(\alpha_{sd}+\beta_{sd})\varepsilon_rk_0^2]}}{\omega^2\mu_0^2(\alpha_{sd}+\beta_{sd})^2}}
\end{align}
\end{figure*}
\begin{figure*}[!t]
\center
\subfloat[]{\label{fig:_Re_er1}
\includegraphics[width=0.9\columnwidth]{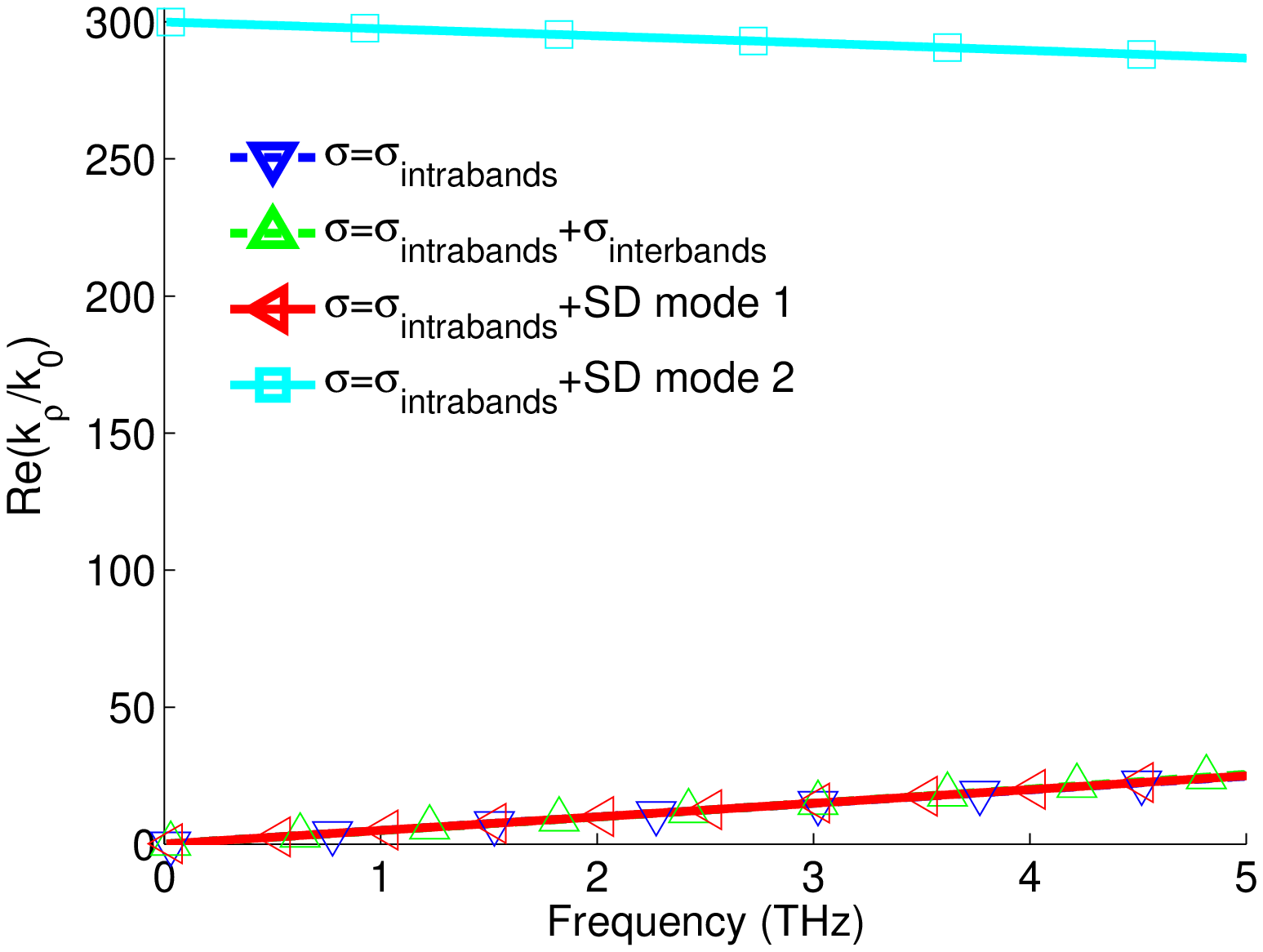}}
\subfloat[]{\label{fig:_Im_er1}
\includegraphics[width=0.9\columnwidth]{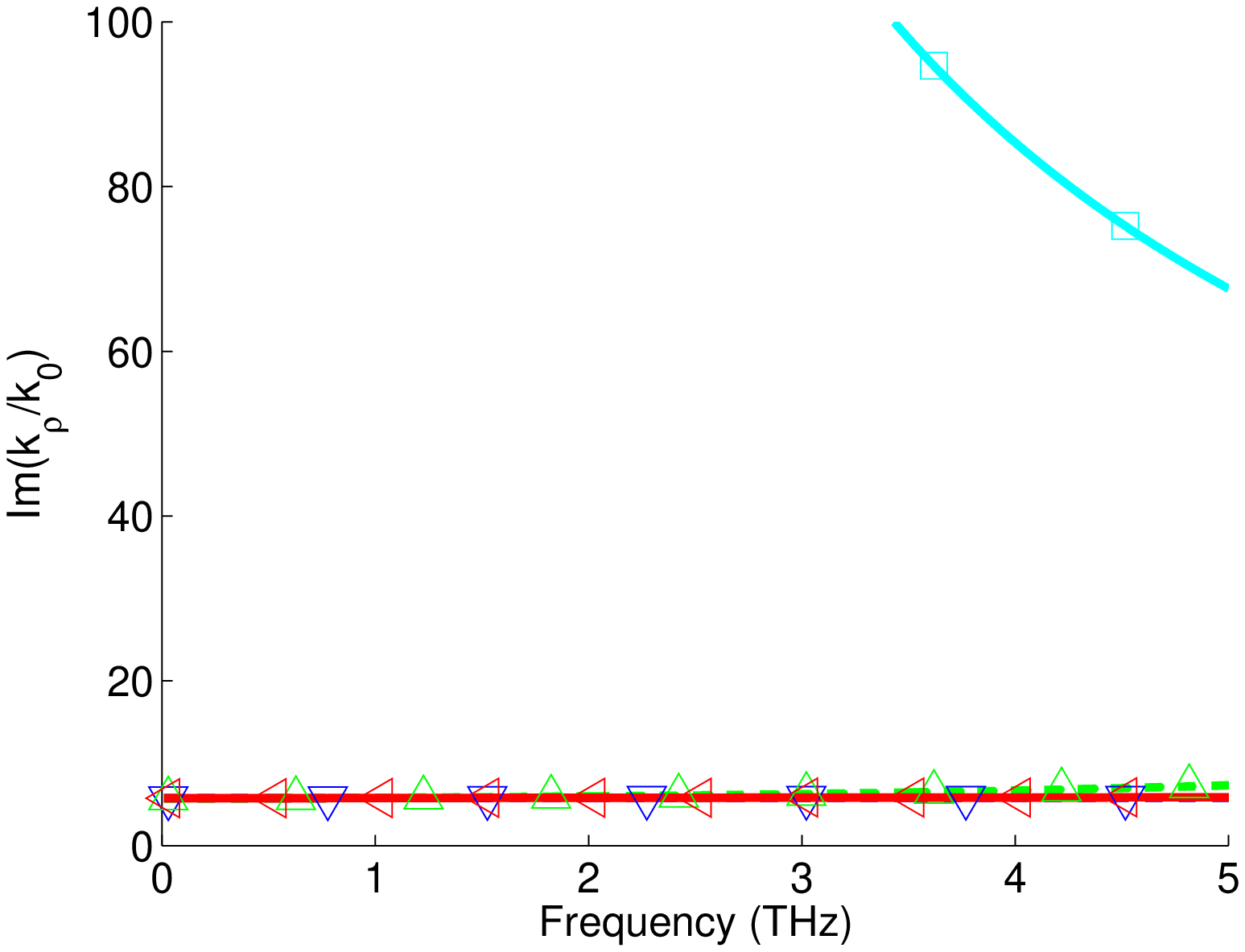}} \\
\subfloat[]{\label{fig:_Re_er11}
\includegraphics[width=0.9\columnwidth]{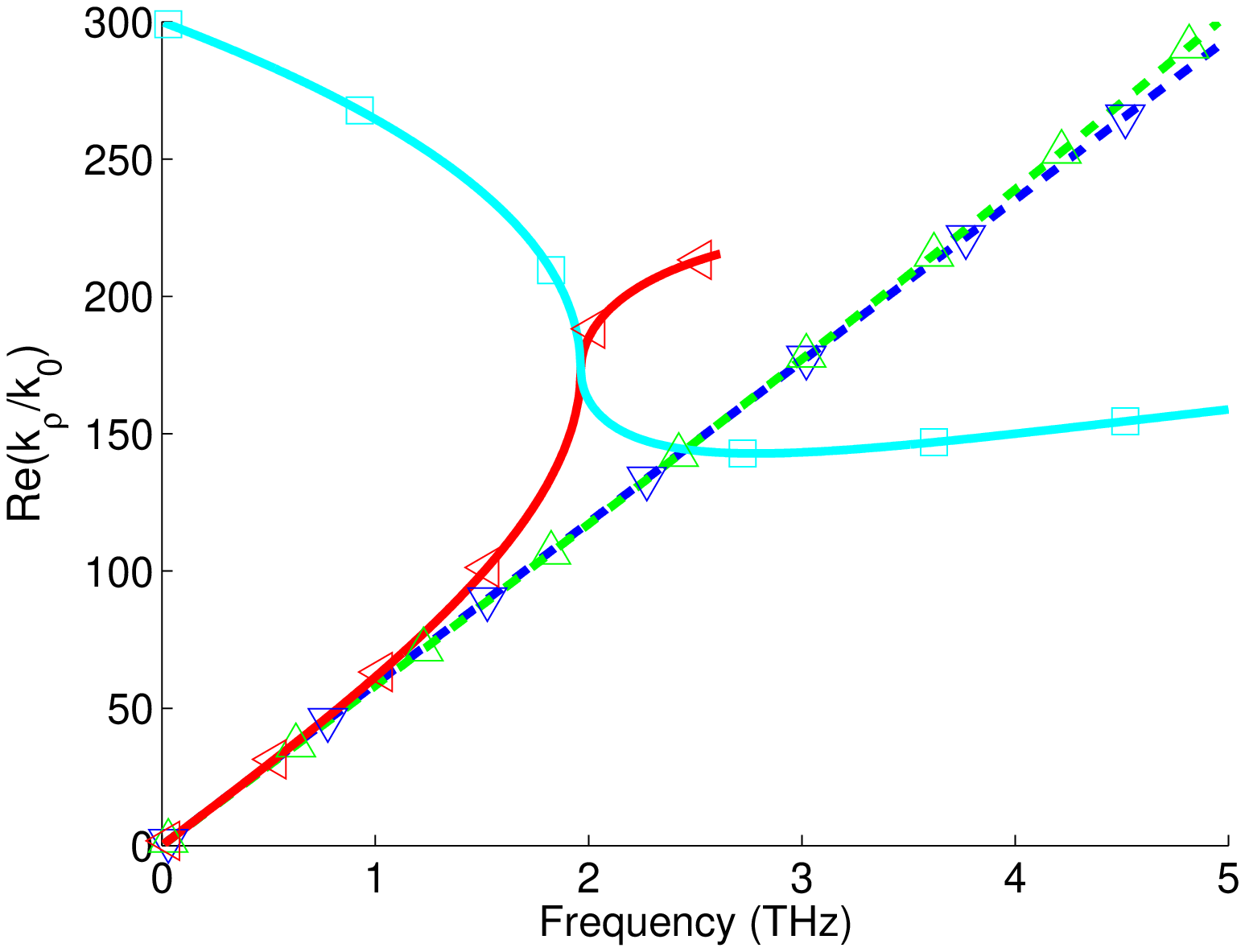}}
\subfloat[]{\label{fig:_Im_er11}
\includegraphics[width=0.9\columnwidth]{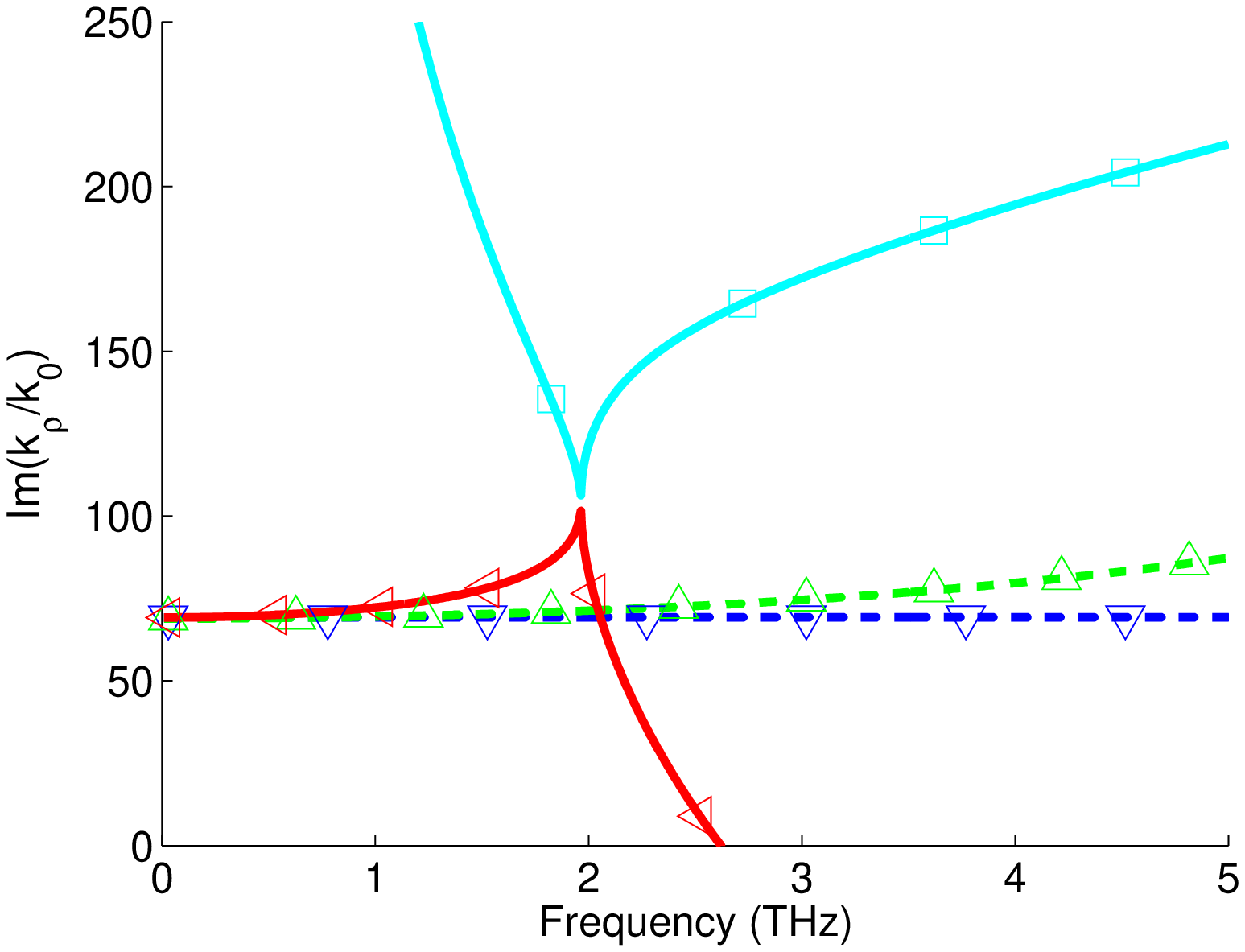}}
\caption{Characteristics of surface waves propagating along a spatially dispersive (SD) graphene sheet versus frequency computed using Eq.~(\ref{Dispersion_relation_TM_homogeneous_medium}). Reference results, related to surface waves propagating along non spatially dispersive graphene \cite{Hanson08} computed taking into account intraband and intraband+interband contributions of conductivity, are included for comparison purposes. (a) and (b) show the normalized propagation constant and losses of surface waves along a free-space standing graphene sheet. (c) and (d) show the normalized propagation constant and losses of surface waves along a graphene sheet embedded into an homogeneous media with $\varepsilon_r=11.9$. Graphene parameters are $\mu_c=0.05$~eV, $\tau=0.135$~ps and $T=300^\circ$~K.}
\label{Fig:_Surface_Waves_Characteristics}
\end{figure*}
\subsection{TE modes}
\label{sec:TE_modes}
The dispersion relation for TE modes propagating along a spatially dispersive graphene sheet can be expressed as

\begin{align}
\frac{\sqrt{\varepsilon_{r1}k_0^2-k_\rho^2}}{\omega\mu_0}+\frac{\sqrt{\varepsilon_{r2}k_0^2-k_\rho^2}}{\omega\mu_0}=-\left[\sigma_{lo}+k_\rho^2(\alpha_{sd}+\beta_{sd})\right].
\label{eq:Dispersion_relation_TE}
\end{align}
In case the graphene sheet is embedded into an homogeneous medium, with $\varepsilon_{r1}=\varepsilon_{r2}=\varepsilon_r$, this equation admits the analytical solution shown in Eq.(\ref{Dispersion_relation_TE_mode}).

Similarly as in the case of TM modes, the cut off frequency of TE modes mainly depends on the local conductivity of graphene, $\sigma_{lo}$. Specifically, an accurate condition for the propagation of these modes along a graphene sheet is $Im(\sigma_{lo})>0$ \cite{Hanson08}.
\section{Numerical Results}
\label{sec:Numerical_results}
\begin{figure*}[!t]
\center
\subfloat[]{\label{fig:_Versus_mu_c}
\includegraphics[width=0.9\columnwidth]{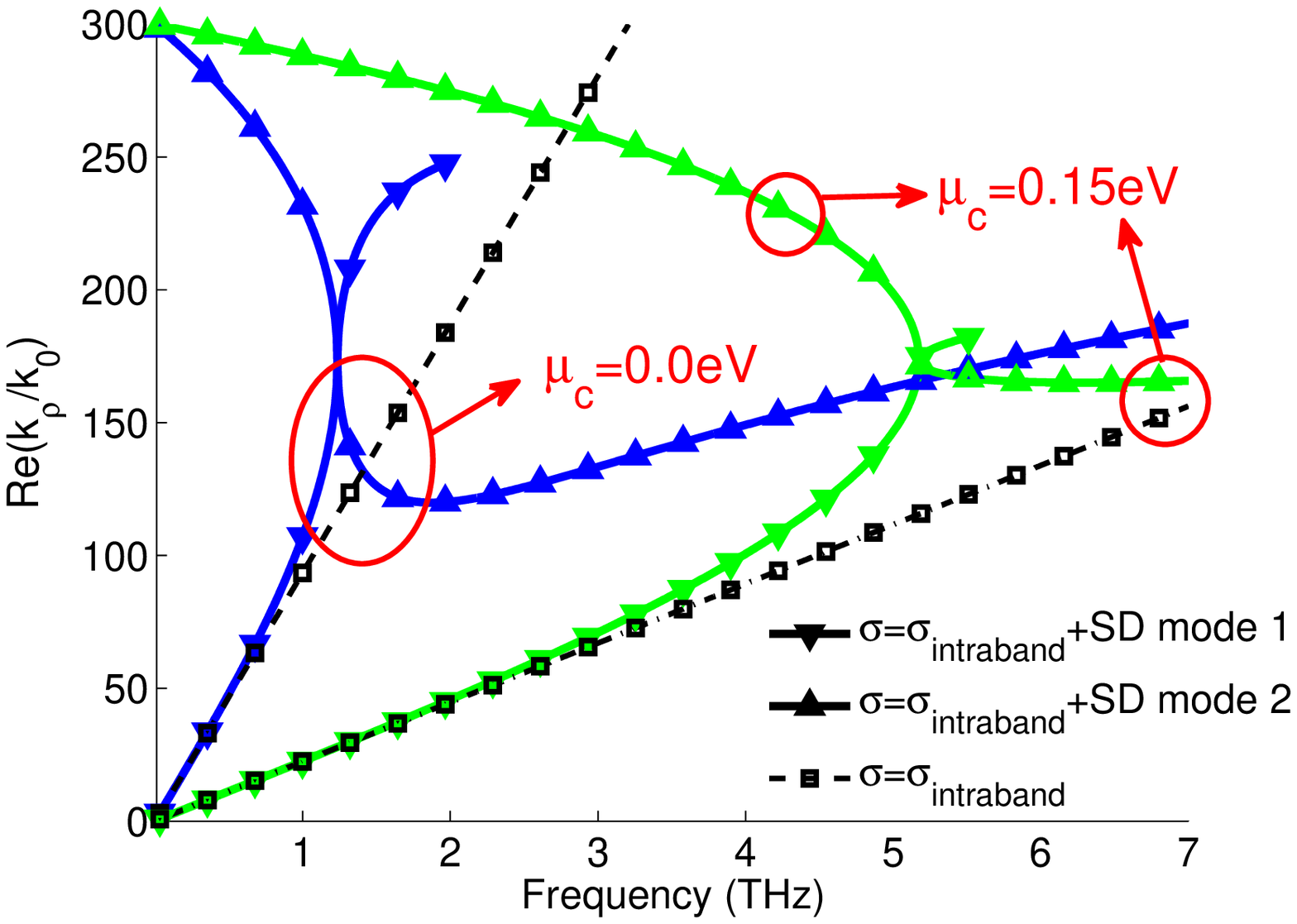}}
\subfloat[]{\label{fig:_Versus_tau}
\includegraphics[width=0.9\columnwidth]{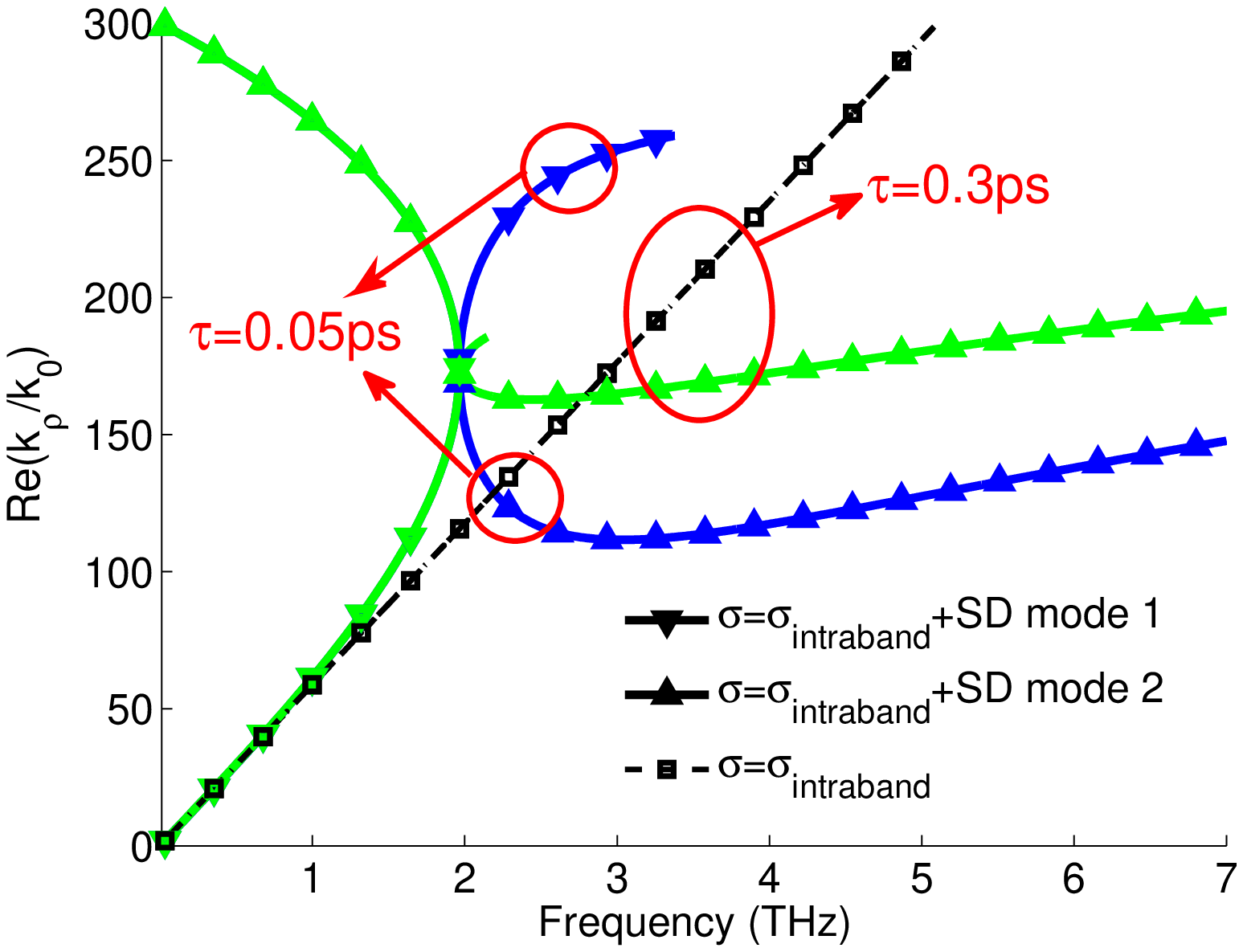}}
\caption{Normalized phase constant of surface waves propagating on a spatially dispersive graphene sheet versus frequency, computed for different values of (a) chemical potential [using $\tau=0.135$~ps] and (b) relaxation time [using $\mu_c=0.05$~eV]. Reference results, related to surface waves on non spatially dispersive graphene sheet computed only taking into account intraband contributions of conductivity \cite{Hanson08}, are included for comparison purposes. The permittivity of the surrounding media is $\varepsilon_r=11.9$ and temperature is $T=300^\circ$~K.}
\label{Fig:_Parametric_study_versus_muc_and_tau}
\end{figure*}
\begin{figure*}[t]
\center
\subfloat[]{\label{fig:_Re_1Thz_vs_er}
\includegraphics[width=0.9\columnwidth]{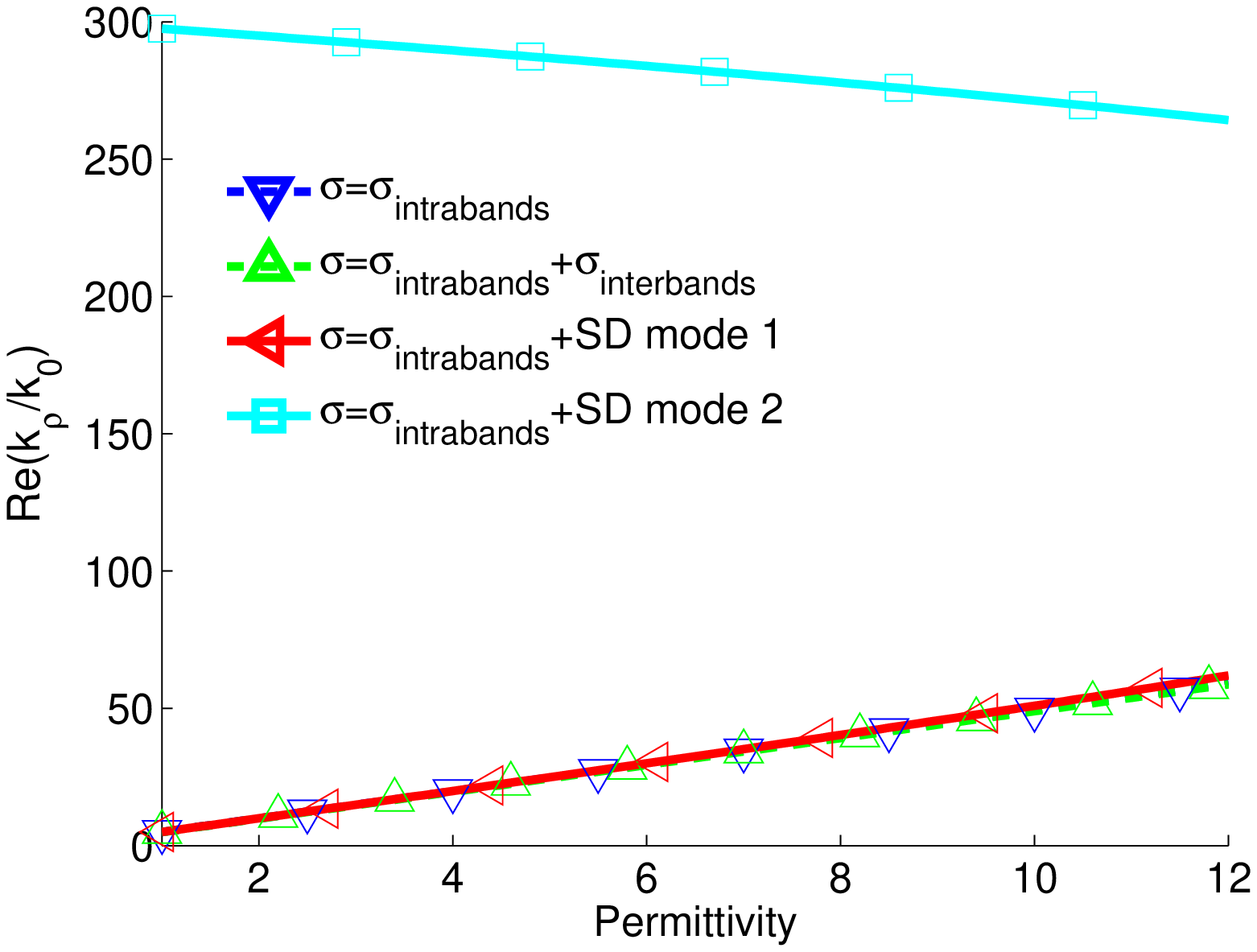}}
\subfloat[]{\label{fig:_Im_1Thz_vs_er}
\includegraphics[width=0.9\columnwidth]{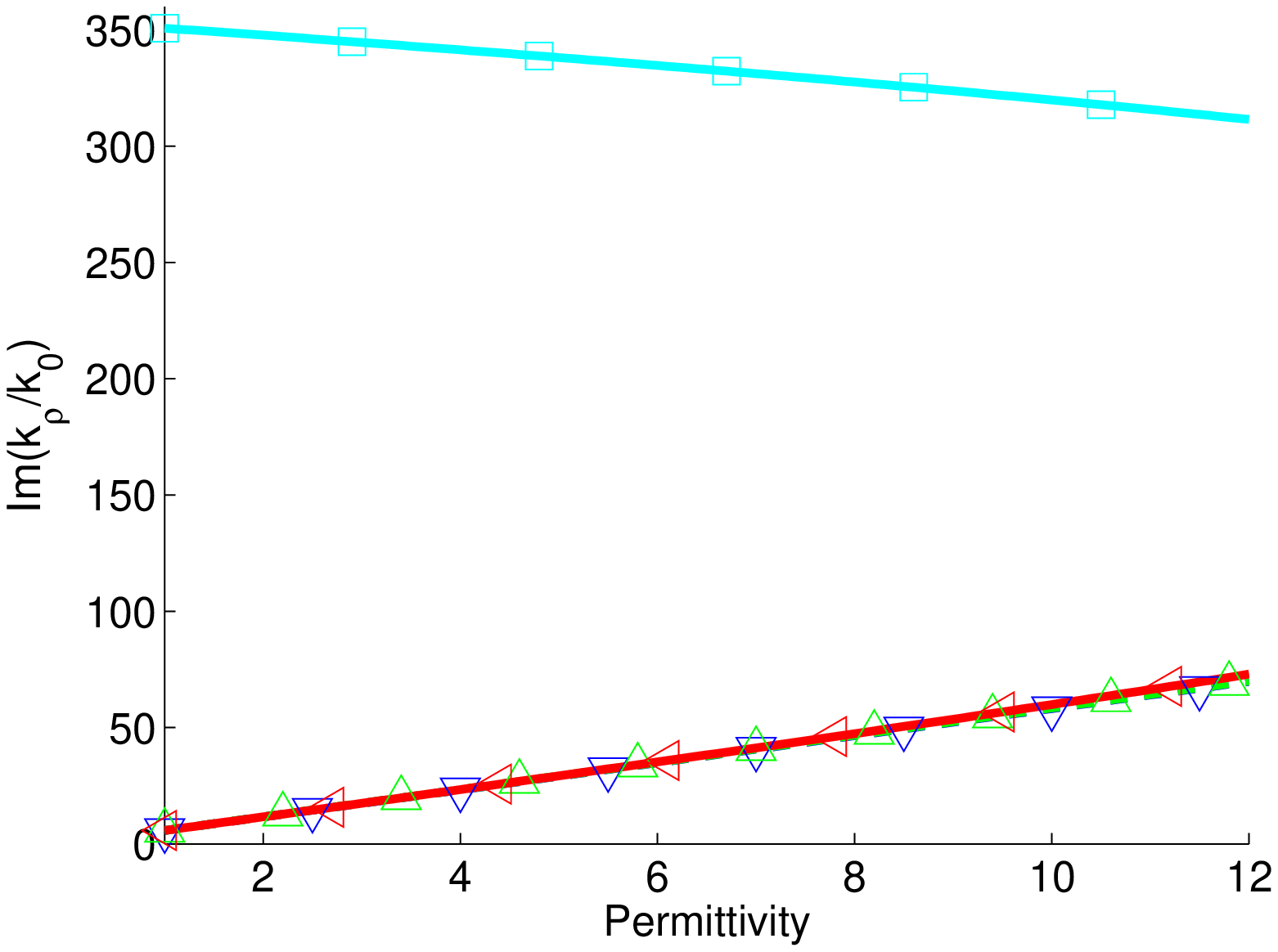}} \\
\subfloat[]{\label{fig:_Re_3Thz_vs_er}
\includegraphics[width=0.9\columnwidth]{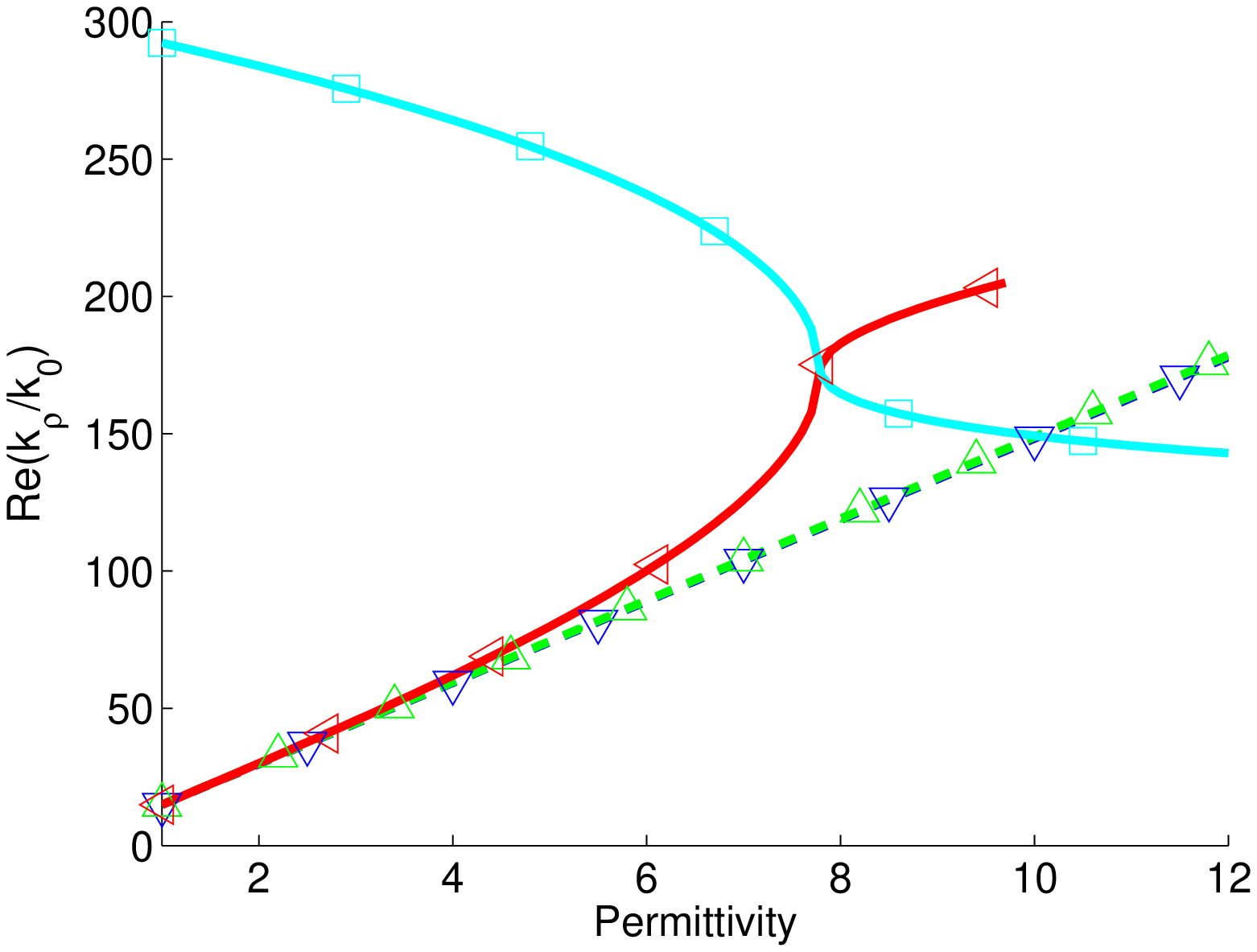}}
\subfloat[]{\label{fig:_Im_3Thz_vs_er}
\includegraphics[width=0.9\columnwidth]{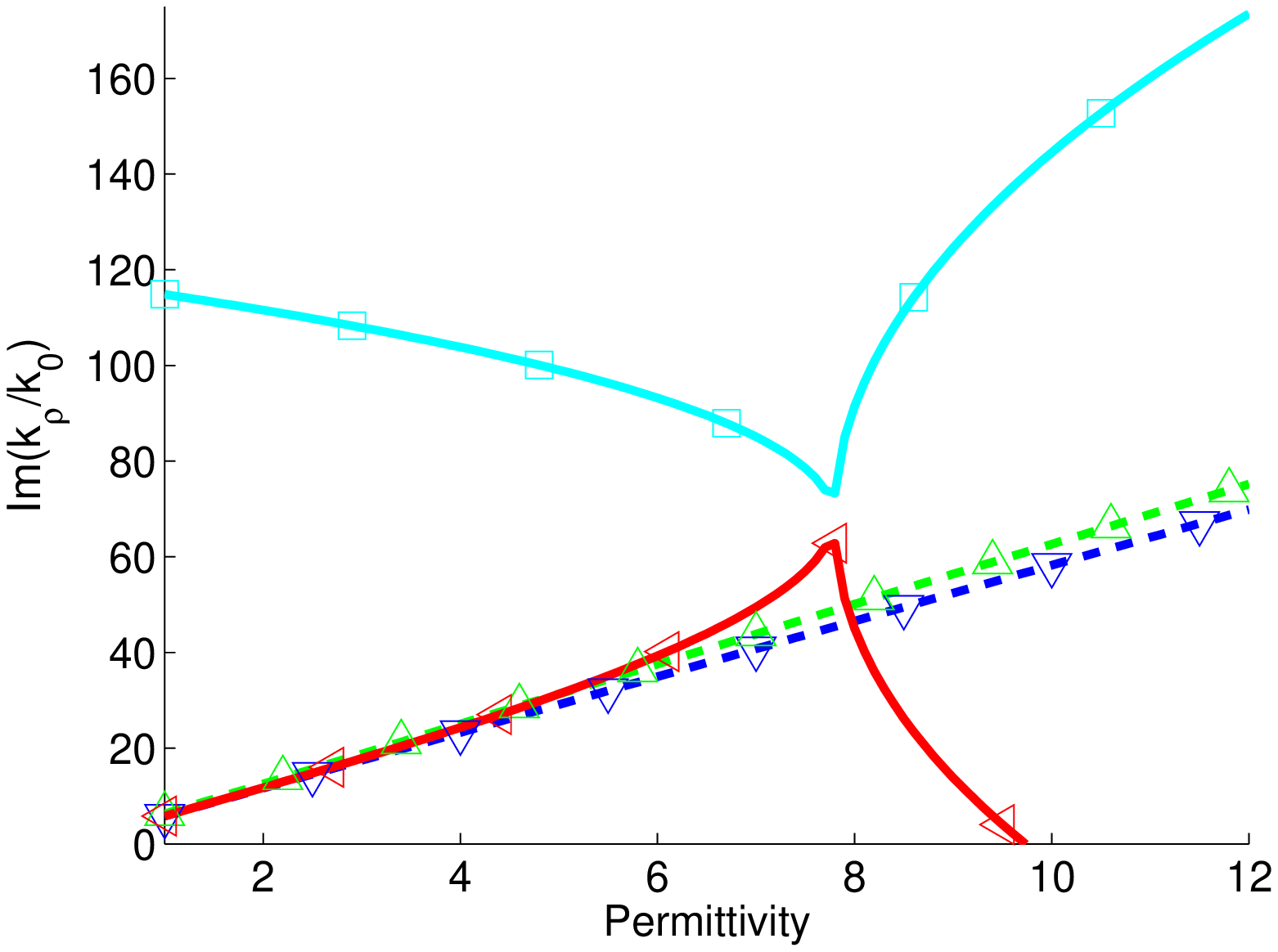}}
\caption{Characteristics of surface waves propagating along a spatially dispersive (SD) graphene sheet as a function of surrounding media permittivity, computed using Eq.~(\ref{Dispersion_relation_TM_homogeneous_medium}). Reference results, related to surface waves propagating along non spatially dispersive graphene \cite{Hanson08} computed taking into account intraband and intraband+interband contributions of conductivity, are included for comparison purposes. (a) and (b) show the normalized propagation constant and losses of the surface waves at $1$~THz. (c) and (d) show the normalized propagation constant and losses of the surface waves at $3$~THz. Graphene parameters are $\mu_c=0.05$~eV, $\tau=0.135$~ps and $T=300^\circ$~K.}
\label{Fig:_Surface_Waves_Characteristics_er_vary}
\end{figure*}
In this section, we investigate the influence of spatial dispersion in the characteristics of surface waves propagating along a graphene sheet, taking into account the surrounding media. Specifically, we study the normalized propagation constant [mode confinement, ${Re}(k_\rho/k_0)$, and losses, $Im(k_\rho/k_0)$] of TM waves along spatially dispersive graphene, and compare the results with the ones obtained neglecting spatial dispersion \cite{Hanson08}. In the numerical study, we focus our results on TM waves, which are known to be of interest for plasmonic devices \cite{Jablan09}, \cite{Barnes03}. Though TE surface waves are also supported by graphene sheets, they present similar characteristics to waves propagating in free space (i.e. $k_\rho\approx k_0$) \cite{Hanson08} and thus have less practical interest. We compute reference results for non spatially dispersive graphene sheet considering intraband and intraband+interband contributions of conductivity. The aim of this comparison is to identify which phenomenon (spatial dispersion or interband contributions) becomes dominant as frequency increases. Furthermore, we will also investigate the influence of spatial dispersion in the propagating surface waves as a function of different parameters of graphene. Our numerical simulations consider graphene at $T=300^{\circ}$~K and a relaxation time of $0.135$~ps, in agreement with measured values \cite{Jablan09, Novoselov04}. The results shown here have been obtained numerically [solving  Eq.~(\ref{eq:Dispersion_relation_TM})] or analytically [from Eq.~(\ref{Dispersion_relation_TM_homogeneous_medium})], depending on the surrounding media of graphene. Besides, note that the analytical solution of Eq. (\ref{eq:Dispersion_relation_explicit}) leads to results with differences smaller than $0.1\%$ with respect to those numerically obtained from Eq.~(\ref{eq:Dispersion_relation_TM}), further confirming the validity of this expression.

First, we consider the case of a graphene sheet with chemical potential $\mu_c=0.05$~eV surrounded by air. Figs. \ref{fig:_Re_er1}-\ref{fig:_Im_er1} show the normalized propagation constant and losses of surface waves propagating along the spatially dispersive sheet. Specifically, two modes are supported by the structure, which are the physical solutions of Eq.~(\ref{Dispersion_relation_TM_homogeneous_medium}). The first mode (denoted as ``SD - mode $1$'') presents extremely similar characteristics as compared to a TM surface mode supported by a non spatially dispersive graphene sheet, computed considering only intraband contributions of graphene. Figs. \ref{fig:_Re_er1}-\ref{fig:_Im_er1} also include similar computations but considering interband contributions as well. The effect of interband contribution become visible at high frequencies, increasing the losses of the mode. Thus, we can conclude that for this particular case the influence of spatial dispersion is negligible in the low THz range. In \cite{Hanson09} similar conclusions were obtained analyzing, in the transformed Fourier domain, the ratio between graphene conductivity and the terms related to spatial dispersion. Interestingly, here we also observe that a second mode, denoted as ``SD - mode $2$'', is supported due to the presence of spatial dispersion. It is observed that this mode is extremely lossy in the band of interest, which greatly limits its possible use in practical applications.

We present in Figs. \ref{fig:_Re_er11}-\ref{fig:_Im_er11} a similar study, but considering now a graphene sheet embedded into an homogeneous material with $\varepsilon_r=11.9$. Results demonstrate that spatial dispersion becomes the dominant mechanism of wave propagation in this case, even at relatively low frequencies, drastically changing the behavior of the two modes supported by the graphene sheet. At low frequencies, the first mode (``SD - mode $1$'') presents very similar characteristics as compared to a TM surface mode on along a graphene sheet neglecting spatial dispersion. However, the confinement and losses of the mode increases with frequency. At around $2$~THz, losses exponentially decreases thus leading to a non-physical improper mode at frequencies above $2.5$~THz. The behavior of the second mode (``SD - mode $2$'') is also affected by spatial dispersion. As frequency increases, the confinement and losses of the mode decreases. Above $2$~THz, the confinement of the mode remains relatively constant and losses slightly increase. It should be noted that although the phase constant of modes $1$ and $2$ intersect at around $2$~THz, their attenuation constants are different at that frequency thus allowing to clearly identify the modes. Besides, note that the frequency where the phase constant of the modes intersect defines the frequency region where the influence of spatial dispersion starts to be dominant. Furthermore, mode coupling is occurring between the two modes supported by the graphene sheet. This coupling is governed by the occurrence of a complex-frequency-plane branch point that migrates across the real frequency axis \cite{Hanson99}, \cite{Yakovlev00}.

Importantly, the influence of spatial dispersion on the characteristics of the supported modes strongly depends on the features of graphene. As an illustration, Fig.~\ref{Fig:_Parametric_study_versus_muc_and_tau} presents the phase constant of the two modes propagating along a graphene sheet (with the parameters employed in Fig.~\ref{fig:_Re_er11}) as a function of the chemical potential $\mu_c$ and the relaxation time $\tau$. We observe in Fig.~\ref{fig:_Versus_mu_c} that the chemical potential controls the frequency where the phase constant of the two modes intersect, which is the frequency where spatial dispersion starts to have a dominant effect on the behavior of the modes. Note that an increase of $\mu_c$ up-shifts this frequency. In addition, Fig.~\ref{fig:_Versus_tau} shows that the relaxation time of graphene determines the phase constant of the surface waves in the frequency region where spatial dispersion is dominant. Increasing $\tau$ modifies the phase constant of the modes, which tend to have a similar behavior versus frequency, and reduces their attenuation losses.
\begin{figure*}[!t]
\center
\subfloat[]{\label{fig:_Re_inhomo}
\includegraphics[width=0.9\columnwidth]{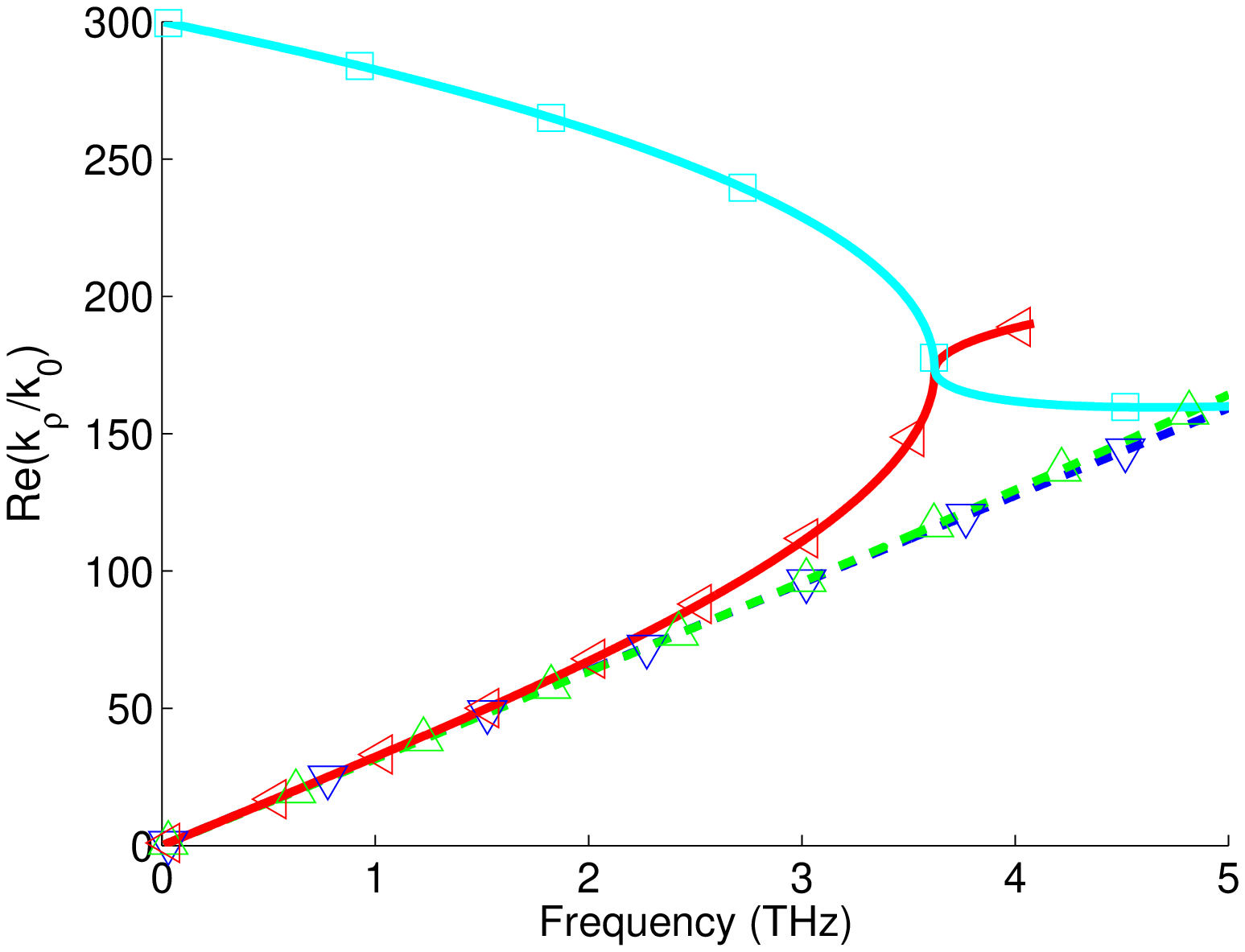}}
\subfloat[]{\label{fig:_Im_inhomo}
\includegraphics[width=0.9\columnwidth]{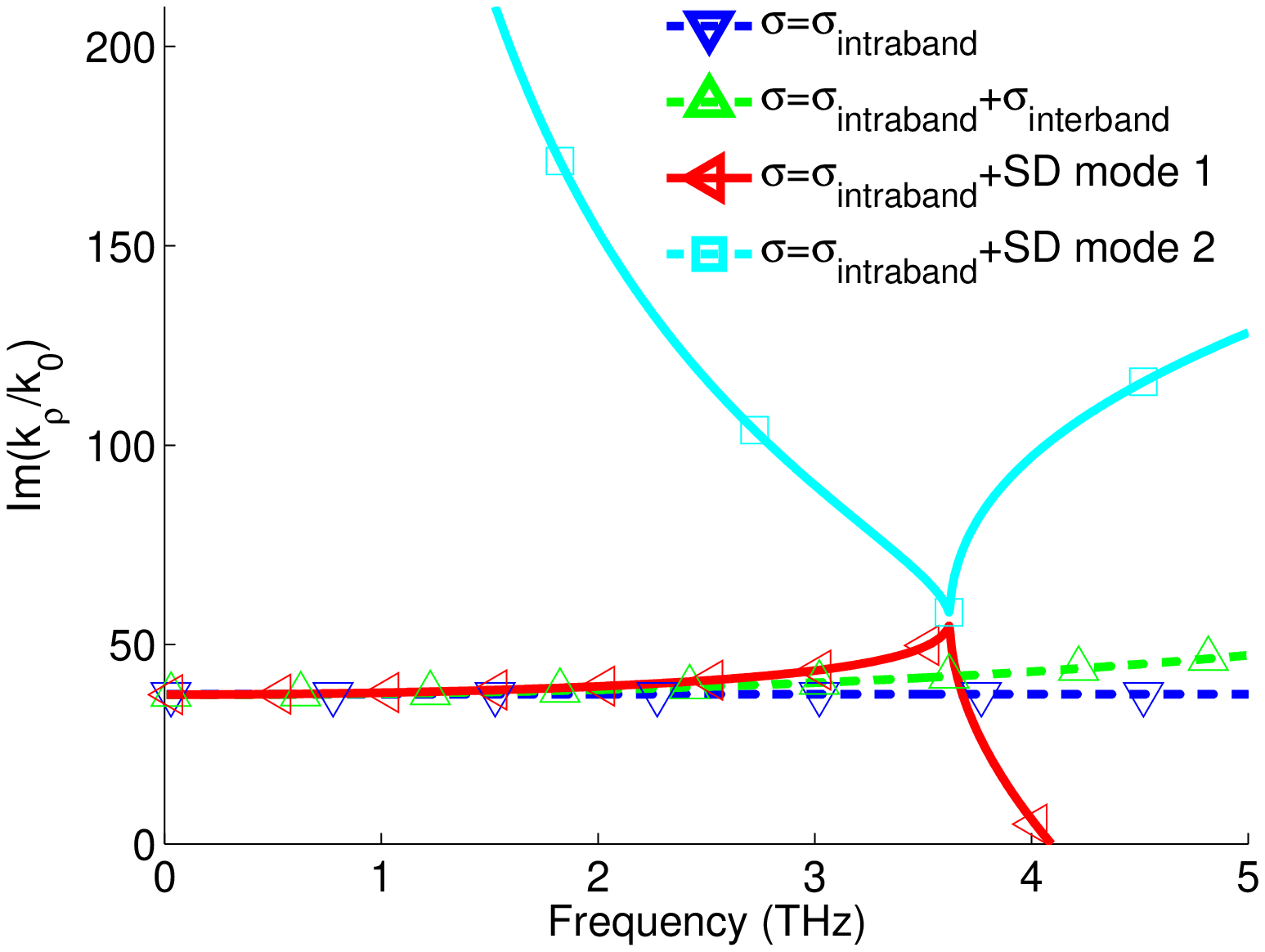}}
\caption{Characteristics of surface waves propagating along a spatially dispersive (SD) graphene sheet deposited on a silicon substrate ($\varepsilon_r=11.9$) versus frequency, computed using Eq.~(\ref{eq:Dispersion_relation_TM}). Reference results, related to surface waves propagating along non spatially dispersive graphene \cite{Hanson08} computed taking into account intraband and intraband+interband contributions of conductivity, are included for comparison purposes. (a) and (b) show the normalized propagation constant and losses of the surface waves, respectively. Graphene parameters are similar to those of Fig.~\ref{Fig:_Surface_Waves_Characteristics}.}
\label{Fig:_Surface_Waves_Characteristics_inhomo}
\end{figure*}

For the sake of completeness, we present in \figref{Fig:_Surface_Waves_Characteristics_er_vary} a study of surface waves propagating along spatially dispersive graphene, but considering now a constant frequency and varying the permittivity of the surrounding media. It is worth mentioning that in practice the material permittivity might affect graphene relaxation time \cite{Dressel02}, but this effect is neglected here for convenience. Figs.~\ref{fig:_Re_1Thz_vs_er}-\ref{fig:_Im_1Thz_vs_er} present the normalized phase and attenuation constants of surface waves propagating along graphene for a fixed frequency of $1$~THz. The first mode is extremely similar to a TM surface mode along graphene neglecting spatial dispersion. As expected from Eq.~(\ref{eq:Dispersion_relation_explicit}), mode confinement and losses increases when the dielectric constant is very high. Interesting, there are many similarities in the behavior of the different modes in this situation and in the previous example, where the surrounding media was constant (with a low permittivity value of $\varepsilon_r=1$) and frequency was increased. Moreover, we present in Figs.~\ref{fig:_Re_3Thz_vs_er}-\ref{fig:_Im_3Thz_vs_er} the same study but at the higher frequency of $3$~THz. It is observed that spatial dispersion becomes the dominant phenomenon for wave propagation. In fact, it leads to propagating modes with similar behavior as in the previous example (see Figs. \ref{fig:_Re_er11}-\ref{fig:_Im_er11}), where we used a fixed large value of permittivity ($\varepsilon_r=11.9$) and varied frequency. This example confirms that, neglecting the frequency dependence of graphene conductivity, permittivity and frequency play a similar role in the characteristics of surface waves propagating along spatially dispersive graphene.

Finally, we investigate in \figref{Fig:_Surface_Waves_Characteristics_inhomo} the propagation of surface waves in a more realistic scenario, which consists of a spatially dispersive graphene sheet deposited on a silicon substrate ($\varepsilon_r=11.9$), as illustrated in \figref{fig:_Draw}. Very similar behavior as compared to the previous examples is obtained. Furthermore, these results demonstrate that the variation of the surrounding media directly controls the frequency range where spatial dispersion is noticeable. Analyzing these variations, we can conclude that increasing the permittivity of the media surrounding graphene down shift the frequencies where the spatial dispersion phenomenon dominates wave propagation.
\section{Conclusions}
\label{sec:Conclusions}
The propagation of surface waves along a spatially dispersive graphene sheet has been addressed. The analysis, that was based on the transverse resonance method extended to handle graphene spatial dispersion, allowed to obtain the desired dispersion relation for surface waves propagating along a graphene sheet. Our results have demonstrated that spatial dispersion is an important mechanism which contributes to the propagation of surface waves along graphene sheets and that its influence cannot be systematically neglected. In fact, the presence of spatial dispersion at frequencies where intraband contributions of graphene dominate lead to surface waves with very different characteristics from those propagating along graphene sheets neglecting spatial dispersion. In addition, the frequency region where spatial dispersion is noticeable depend on the surrounding media and the chemical potential of graphene, while the features of the propagating surfaces waves are mainly determined by the relaxation time. These results demonstrate that the influence of spatial dispersion on surface waves propagating on graphene sheets should be rigorously taken into account for the development of novel plasmonic devices in the low THz range.

The study presented here has been based on a theoretical model of graphene which characterizes it as an infinitesimally thin layer with an associated tensor conductivity. This model uses the relaxation time approximation, only takes into account intraband contributions of graphene and is valid in the absence of external magnetostatic biasing fields. Further work is needed in this area to analyze the behavior of surface waves along spatially dispersive graphene when these assumptions are not satisfied.
\section*{\label{Acknoled} Acknowledgement}
This work was supported by the Swiss National Science Foundation (SNSF) under grant $133583$ and by the EU FP$7$ Marie-Curie IEF grant ``Marconi", with ref. $300966$. The authors wish to thank Prof. G.~W.~Hanson (University of Wisconsin-Milwaukee, USA), Dr. Garcia-Vigueras and Dr. E. Sorolla-Rosario (\'Ecole Polytechnique F\'ed\'erale de Lausanne, Switzerland) for fruitful discussions.

\end{document}